\documentclass[twocolumn,pra,superscriptaddress,preprintnumbers]{revtex4}
\usepackage{mathrsfs,graphicx,amssymb,amsbsy}

\def\scr{\mathscr}

\def\SU{{\scr U}}

\def\avg#1{\langle#1\rangle}    \def\<{\langle}     \def\>{\rangle}

\def\pt{\partial}
           
\def\sig{\sigma}    \def\del{\delta}    \def\Del{\Delta}
\def\eps{\epsilon}    
\def\up{\uparrow}   \def\down{\downarrow}

\def\Bk{{\mathbf k}}  
    \def\BK{{\mathbf K}} \def\Bl{{\mathbf l}}

\def\Bx{{\mathbf x}}  \def\B0{{\mathbf 0}}

\def\be{\begin{equation}}   \def\ee{\end{equation}}
\def\bea{\begin{eqnarray}}  \def\eea{\end{eqnarray}}
\def\nn{\nonumber}

 \def\Uch{U(1)$_\mathrm{charge}$}
\def\td#1{\tilde{#1}}

\begin{document}
\preprint{MIT-CTP-3437}

\title{Spin-Dependent Hubbard Model and a Quantum
Phase Transition in Cold Atoms}

\author{W. Vincent Liu}
\author{Frank Wilczek}
\affiliation{Center for Theoretical Physics, Department of Physics,
Massachusetts Institute of Technology,
Cambridge, Massachusetts 02139}
\author{Peter Zoller}
\affiliation{Institut f\"ur Theoretische Physik, Universit\"at Innsbruck,
  A--6020 Innsbruck, Austria}

\date{\today}

\begin{abstract}

We describe an experimental protocol for introducing spin-dependent
lattice structure in a cold atomic fermi gas
using lasers.  It can be used to realize Hubbard
models whose hopping parameters depend on spin and whose
interaction strength can be controlled with an external
magnetic field.  We suggest that exotic
superfluidities will arise in this framework.
An especially interesting possibility is a class
of states that support coexisting superfluid and
normal components, even at zero temperature.
The quantity of normal component varies with
external parameters.  We discuss some aspects of
the quantum phase transition that arises at the point where it vanishes.

\end{abstract}

\maketitle

%\thispagestyle{empty}
%\pagebreak

%%%%%% Introduction

\section{Introduction}

Cold atom systems can be used to explore important problems of
condensed matter physics in new ways. For example, recent rapid
development of the ultracold atomic gas in optical
lattices~\cite{UCA:nature,Cirac+Zoller-PhysicsToday:04} supported the
observation of the superfluid-Mott insulator transition in cold atomic
bosons confined in an optical
lattice~\cite{Bloch-SF-Mott:02,Orzel+:01,Bloch-CCEntanglement:03,Stoeferle+Esslinger-Mott1D:04}.
In a wider context, methods of ``engineering'' various lattice model
systems with bosonic and fermionic atoms have been proposed
\cite{Jaksch+BruderETAL-ColdBosonicAtomOpticalLattices:98,Jaksch+ETAL-EntanglementColdCollision:99,Lewenstein+SantosETAL-AtomBoseMixtOpticalLattices:04,Damski+ZakrzewskiETAL-AtomBoseAndeGlas:03,Kuklov++Duan:03,Paredes+Cirac-CooperPairsLuttingerOpticalLattice:03,Ho+CazalillaETAL-OptcaliLattice1D2D:04,Pachos-Rico:04pre,Hofstetter+:02,Rabl+DaleyETAL-DefectSuppressed:03},
opening prospects for exploring exotic new phases. In recent years
achieving superfluidity in cold atomic {\em fermions} has become a
major goal, involving several experimental programs
~\cite{Hulet:03,Greiner+Regal+Jin:03,Cubizolles+Salomon-MolecularBECLi:03,
Zwierlein+KetterleETAL-LiMolecularBEC:03,Regal+Greiner+Jin-KCrossover:04,Bartenstein+Grimm-LiCrossover:04,Zwierlein+Ketterle-LiCrossover:04,Thomas:04,Ferlaino+Inguscio-ExpansionFermiBose:04}.
Here we consider techniques that permit one to introduce both sorts of
complexity in one system, with controlled band structures and
interactions that depend on spin.  They involve counterpropagating
laser beams, that together generate a standing light wave which leads
to different AC Stark shifts for the spin up and spin down components
of alkali atoms (such as $^{40}$K) in their ground state
~\cite{Cirac+Zoller-PhysicsToday:04,Jaksch+ETAL-EntanglementColdCollision:99}.
We then discuss an interesting new phase of matter that might arise in
this context, involving coexistence of normal Fermi liquid and
superfluid components
~\cite{Liu-Wilczek:03,Wu-Yip:03,Gubankova+:03,Liu-Wilczek:03:comm,Bedaque:03,Liao:03pre,Deb:03pre},
and the quantum phase transition between this state and conventional
Bardeen-Cooper-Schrieffer~\cite{BCS:57} (BCS) superfluidity.

\section{Spin-dependent optical lattices for cold fermionic atoms}
\label{sec:spindeplattice}

In this section we describe how to realize, using cold atoms in an
optical lattice, a rather general band Hubbard-type model, with
separate, tunable effective masses and filling factors for the two
spins
\cite{Jaksch+BruderETAL-ColdBosonicAtomOpticalLattices:98,Jaksch+ETAL-EntanglementColdCollision:99,
Hofstetter+:02}.

\subsection{The setup of an optical lattice model}

Atoms in an off-resonant laser field exhibit a second order AC Stark shift
of their ground state levels. This shift is proportional to the light
intensity. The intensity may vary in space, for example, by forming a
standing light wave from counterpropagating laser beams. For the
center-of-mass motion of the atoms, spatially varying AC Stark shifts play
the role of a conservative potential. In particular, a standing light wave
leads to a periodic intensity pattern which results in a periodic potential,
i.e. an optical lattice. By superimposing lattice beams from different
directions and different intensity an effectively 1D, 2D or 3D lattice can
be built.

\begin{figure}[htbp]
\begin{center}
\includegraphics[width=\linewidth]{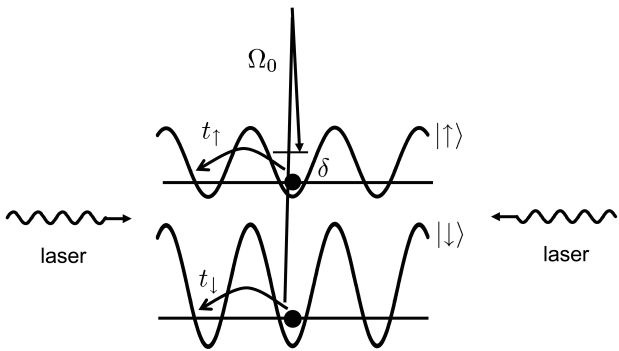}
\end{center}
\caption{Spin-dependent lattice: two counterpropagating laser beams generate
a standing light wave which leads to different AC Stark shifts for the spin
up and spin down components of the ground state atoms ($\protect\sigma =
\uparrow, \downarrow$). As a result the tunneling matrix elements $%
t_\protect\protect\sigma$ for the two spin components are different. The
spin up and down components are coupled by a Raman laser with Rabi frequency
$\Omega_0$ and detuning $\protect\delta$.}
\label{fig:pz1}
\end{figure}

Consider a cloud of cold fermionic atoms in 3D optical lattice
\cite{Hofstetter+:02}. We assume that only two of the internal
ground states participate in the dynamics, which we call spin up
and down, $|\uparrow\rangle$ and $|\downarrow\rangle$. The
corresponding effective Hamiltonian is
\begin{eqnarray}
H &=&\sum_{\sigma }\int d^{3}x~\psi _{\sigma }^{\dag }(\Bx)\left[ -\frac{%
\hbar ^{2}}{2m}\nabla ^{2}+V_{\sigma }(\Bx)\right]
\psi _{\sigma}(\Bx) \nn \\
&&-\delta \int d^{3}x~\psi _{\uparrow }^{\dag
}(\Bx)\psi_{\uparrow} (\Bx) \nn \\
%%&&+\frac{1}{2}\int d^{3}x~\psi _{\uparrow }^{\dag }(\Bx)
%%\Omega_{1}(\Bx)\psi _{\downarrow }(\Bx)+\text{h.c.} \nn \\
&&+\frac{\Omega_0}{2}\int d^{3}x~\psi _{\uparrow }^{\dag }(\Bx)
\psi _{\downarrow }(\Bx)+\text{h.c.} \nn \\
&&+\frac{4\pi a_{s}\hbar ^{2}}{m}\int d^{3}x
\psi _{\uparrow }^{\dag }(\Bx)
\psi _{\downarrow }^{\dag }(\Bx)\psi _{\downarrow }(\Bx)\psi
_{\uparrow }(\Bx) \label{Hamil}
\end{eqnarray}
with $\psi_{\sigma=\uparrow,\downarrow}(\Bx)$ fermionic field operators
obeying the usual anticommutation relations. The various terms contributing
to the Hamiltonian are illustrated in Fig.~\ref{fig:pz1}.

The first line in (\ref{Hamil}) contains the the kinetic energy and the
optical lattice potential $V_{\sigma}(\Bx)$ generated by the laser
beams, obtained in second order perturbation theory for coupling of the
ground state levels to the excited atomic states. In the simplest case of
three orthogonal laser beams this potential has the form
\begin{equation}
V_{\sigma}(\Bx)=\sum_{\ell=1}^{3}V_{\sigma\ell}^{(0)}\sin^{2}kx_{\ell }.
\label{3Dlattice}
\end{equation}
Here $k=2\pi/\lambda$ is the wave vector of the light, $V_{\sigma\ell}^{(0)}$
is proportional to the lattice beam intensity in the direction $\ell$ and
the dynamic atomic polarizability at the laser frequency of the level (spin
state) $\sigma$.

Note that in Eq. (\ref{Hamil}) we have ignored \emph{cross} AC
Stark terms coupling two different spin components. These are
negligible for the situation indicated in Fig.~\ref{fig:pz2},
where the AC Stark shifts are much smaller than the (bare) energy
of the atomic levels, which can be achieved, e.g., by applying a
magnetic field to split the magnetic sublevels of the atoms.
Furthermore, we have assumed in (\ref{3Dlattice}) that the AC
Stark shifts of lattice beams in different directions can be added.
Depending on the laser configuration there may be interference
terms. However, for lattice beams in different directions with
(slightly) shifted optical frequencies (obtained by modulating the
lattice beams), these interference terms will average to zero. We
have also ignored contributions from spontaneous emission.   These
can, for sufficiently large detunings of the lattice beams from
the excited states, be made arbitrarily small.  Indeed, for large
detuning $\Delta\gg\Gamma$ from excited states with width
$\Gamma$, the spontaneous emission rate scales as
$\Gamma/\Delta^{2}$ while the lattice potential scales as
$1/\Delta$, so spontaneous emission becomes, on the relevant
dynamical timescale, negligible.

A key element in (\ref{Hamil}) is the assumption
that the lattice potential $V_{\sigma}(\Bx)$ is spin dependent, as
illustrated graphically in Fig.~\ref{fig:pz1}.
Immediately below we will discuss in
detail specific atomic and laser configurations
that allow us, by varying laser parameters, to
engineer this spin dependence.

The second line in Eq. (\ref{Hamil}) describes the coupling of the two spin
states via a Raman transition with effective two-photon Rabi frequency $%
\Omega_{0}$, and Raman detuning $\delta$. The Hamiltonian has been
written in the  frame, where the optical frequencies have been
transformed away, so that only the detuning $\delta$ appears in
(\ref{Hamil}).

Finally, the last term in (\ref{Hamil}) is the usual atom-atom
interaction described by a pseudo potential with scattering length
$a_{s}$. In atomic experiments the sign and magnitude of the
scattering length can be manipulated, for example, with an
external magnetic field via Feshbach resonances
\cite{Regal+Greiner+Jin-KCrossover:04,Bartenstein+Grimm-LiCrossover:04,Zwierlein+Ketterle-LiCrossover:04}.The
derivation of (\ref{Hamil}) assumes explicitly that the scattering
length is much smaller than the lattice spacing (given by
$\lambda/2$).   In more general circumstances, the interaction
will extend beyond nearest neighbors.

For single atoms the energy eigenstates in the optical lattice are
conveniently described by a band structure and Bloch wave
functions. An appropriate superposition of the Bloch wave
functions forms a set of Wannier states which are localized at the
various lattices sites. For typical experimental parameters the
frequencies associated with dynamics of cold atoms in the lattice
are much smaller than the excitation energies in the optical
potential (excitation to higher Bloch bands). Expanding the field
operators in the set of Wannier functions of the lowest Bloch
band, $\psi_{\sigma}(\Bx)=\sum_{i}w_{\sigma }(\Bx -
\Bx_{i})a_{\sigma i}$, we obtain the Hubbard (single band)
Hamiltonian
\cite{Jaksch+BruderETAL-ColdBosonicAtomOpticalLattices:98}
\begin{eqnarray}
H & =&-\sum_{\sigma\langle i,j\rangle} t_{\sigma}\left(  c_{\sigma
i}^{\dagger }c_{\sigma j}+\mathrm{h.c.}\right)
  +h \sum_{i}\left(c_{\uparrow i}^{\dagger}c_{\uparrow i}
- c_{\down i}^{\dagger}c_{\down i}  \right)  \nn \\
&& +\frac{\Omega_{0}}{2}\sum_{i}\left( c_{\uparrow
i}^{\dagger}c_{\downarrow i}+\mathrm{h.c.}\right)
-U\sum_{i}c_{\uparrow i}^{\dagger}c_{\downarrow
i}^{\dagger}c_{\downarrow i}c_{\uparrow i}\,.
\label{eq:HubbardH:atom}   %%\label{HamilHubbardlattice}
\end{eqnarray}
%% \begin{eqnarray}
%% H & =&-\sum_{\sigma\langle i,j\rangle}T_{\sigma}\left( a_{\sigma i}^{\dagger
%% }a_{\sigma j}+\mathrm{h.c.}\right)  \label{HamilHubbardlattice} \\
%% && -\delta\sum_{i}a_{\uparrow i}^{\dagger}a_{\uparrow i} +\frac{1}{2}%
%% \Omega_{1}\sum_{i}\left( a_{\uparrow i}^{\dagger}a_{\downarrow i}+\mathrm{%
%% h.c.}\right)  \nonumber \\
%% &&-U\sum_{i}a_{\uparrow i}^{\dagger}a_{\downarrow i}^{\dagger}a_{\downarrow
%% i}a_{\uparrow i}  \nonumber
%% \end{eqnarray}
In the Hamiltonian (\ref{eq:HubbardH:atom}),
$h=-\delta/2$ plays the role of ``magnetic field'', $t_{\sigma}$
is a spin dependent hopping term which follows from
the spin dependent optical potential, and $U$ is an onsite
interaction. The dependence of the hopping amplitude $t_{\sigma}$
and onsite interaction $U$
on the optical lattice parameters is given by $t_{\sigma}=E_{R}(2/\sqrt{\pi}%
)(V_{0\sigma}/E_{R})^{3/4}\exp(-2(V_{0\sigma}/E_{R})^{1/2})$ and $%
U=E_{R}ka_{s}\sqrt{8/\pi}(V_{0\sigma }/E_{R})^{3/4}$ with $%
E_{R}=\hbar^{2}k^{2}/2m$ the recoil frequency of the atoms, and $%
k=2\pi/\lambda$ the wave vector of the light. Thus the ratio of tunneling to
onsite interaction can be controlled via the depth of the optical lattice.

\subsection{Tuning spin-dependent tunneling}

\begin{figure}[htbp]
\begin{center}
\includegraphics[width=\linewidth]{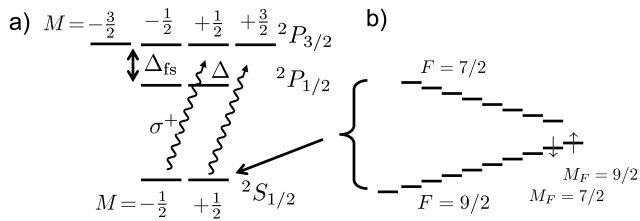}
\end{center}
\caption{a) Atomic level scheme of Alkali atoms. A $\protect\sigma^+$
polarized lattice beam is tuned between the two excited state fine structure
components resulting in different AC Stark shifts for the ground state
levels as explained in the context of
Eq.~({\protect\ref{SpinDepACStarkShift}}).
b) Two possible ``spin up" and ``spin down" states are illustrated the case of
$^{40}$K atoms. }
\label{fig:pz2}
\end{figure}

For alkali atoms, i.e. the atoms used in present cold fermi gas
experiments, the difficulty in obtaining a \emph{spin dependent}
lattice arises from the $s$-wave character of the ground state. It
implies that  the AC Stark shift induced by far off-resonant
driving fields is the same for all (hyperfine) ground state
levels. Fortunately, however, heavy alkali atoms such as $^{40}K$,
have a large fine structure splitting of the first excited state
\cite{Jin:99}.  This permits us to obtain a spin dependent optical
potential by tuning the lattice lasers between the fine structure
levels,  still remaining far off resonance to suppress spontaneous
emission \cite{Jaksch+ETAL-EntanglementColdCollision:99}. As
illustrated in Fig.~\ref{fig:pz2}, the ground
states $n\;^{2}S_{1/2}~M=\pm1/2$ states couple to excited states $%
n\;^{2}P_{3/2,1/2}$ with right
(respectively, left) circularly polarized light
$\sigma^±$
according to the selection rules $\Delta M=\pm1$.   This gives rise to a spin
dependent AC Stark shift.

For example, the AC Stark shift of the two ground
states in circularly polarized light with amplitude $\mathcal{E}$ is
\begin{eqnarray}
\Delta E_{^{2}S_{1/2}M=+1/2}(\sigma^{+}) & =\frac{|\mu_{3/2} \mathcal{E}|^{2}%
}{\hbar(\Delta-\Delta_{\mathrm{fs}})}  \label{SpinDepACStarkShift} \\
\Delta E_{^{2}S_{1/2}M=-1/2}(\sigma^{+}) & =\frac{|\mu_{1/2}\mathcal{E}|^{2}%
}{\hbar\Delta}+\frac{|\mu_{3/2}\mathcal{E}|^{2}}{\hbar(\Delta-\Delta_{%
\mathrm{fs}})}\, ,  \nonumber
\end{eqnarray}
keeping only the dominant contributions from the lowest lying
excited states.  Here the detuning from the $^{2}P_{3/2}$ state is denoted
$\Delta$, $\Delta_{\mathrm{fs}}$ is the fine structure splitting, and the
dipole matrix elements for the transitions from the ground state are $%
\mu_{3/2,1/2}$. We see that for detunings between the fine structure states $%
0 < \Delta < \Delta_{\mathrm{fs}}$ the AC Stark shift $\Delta
E_{^{2}S_{1/2}M=-1/2}$ switches from positive to negative values, while $%
\Delta E_{^{2}S_{1/2}M=+1/2}$ is always negative, giving rise to a strongly
spin dependent lattice potential.  We can vary the strength of the Stark
shift by varying laser power.

Thus, the laser
configuration consisting of unbalanced right and left circularly
polarized polarization components gives rise to a spin dependent
lattice, where the resulting AC Stark shifts are a sum of the shifts in the
first and second line of (\ref{SpinDepACStarkShift}), weighted according to
their fraction of $\sigma^{+}$ and $\sigma^{-}$ component. For linearly
polarized light the AC Stark shifts of the two states are equal.

It is easy to include hyperfine splitting of the atomic ground
states~\cite{Jaksch+ETAL-EntanglementColdCollision:99}. Consider
an atomic ground state electron which is coupled to a nuclear spin
$I$. The resulting total angular momentum
is $F=|I\pm 1/2|$, and the hyperfine ground states have the form $%
|F,M_{F}\rangle =a|I,M_{F}+1/2\rangle |^{2}S_{1/2}M=-1/2\rangle
+b|I,M_{F}-1/2\rangle |^{2}S_{1/2}M=+1/2\rangle $ where $a$ and
$b$ are Clebsch-Gordan coefficients.  Assuming that the Zeeman
hyperfine states are split by a constant magnetic field (compare
the discussion following (\ref{Hamil})),  the AC Stark shift of this
state will be the weighted sum $\Delta E_{FM_{F}}=|a|^{2}\Delta
E_{^{2}S_{1/2}M=-1/2}+|b|^{2}\Delta E_{^{2}S_{1/2}M=+1/2}$. Thus a
pair of hyperfine ground state levels will have a spin dependent
lattice provided that the Clebsch-Gordan coefficients in the
decomposition are sufficiently different. These spin dependent
lattices were first proposed for quantum computing purposes by
Jaksch \emph{et al.}
\cite{Jaksch+ETAL-EntanglementColdCollision:99} and recently
implemented in experiments with cold Rb atoms by Bloch and
collaborators~\cite{Bloch-CCEntanglement:03}.

\begin{figure}[htbp]
\begin{center}
\includegraphics[width=\linewidth]{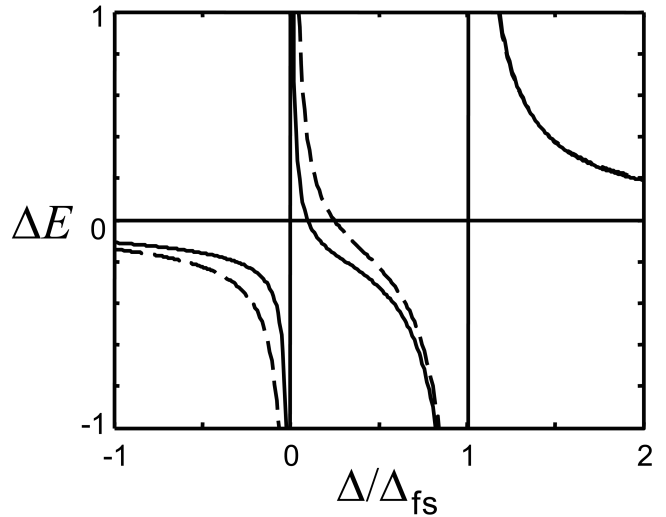}
\end{center}
\caption{AC Starkshift (in arbitrary units) of the $^{40}$K hyperfine
   $M_F=9/2$ (solid line) and $M_F=7/2$ (dashed line) states (compare
   Fig.~\protect\ref{fig:pz2}b) as a function of the detuning
   $\Delta$. The value of $\Delta=0$ corresponds to the $^{2}$P$%
   _{1/2}$excited fine structure state, while
   $\Delta=\Delta_{\mathrm{fs}}$ is the $^{2}$P$_{3/2}$ resonance. We
   note the strong spin dependence (i.e.~dependence on the internal
   state)  for detuning between the two fine structure states. }
\label{fig:pz3}
\end{figure}

\begin{figure}[htbp]
\begin{center}
\includegraphics[width=\linewidth]{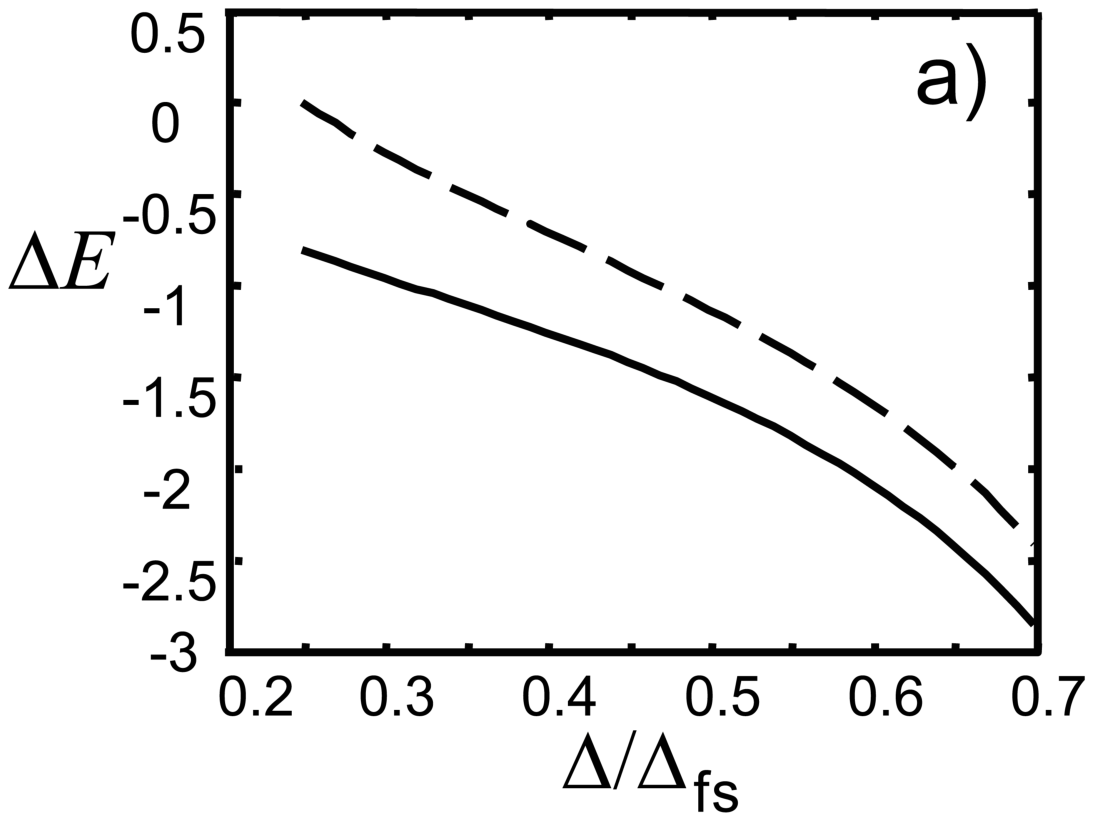}\\
\includegraphics[width=\linewidth]{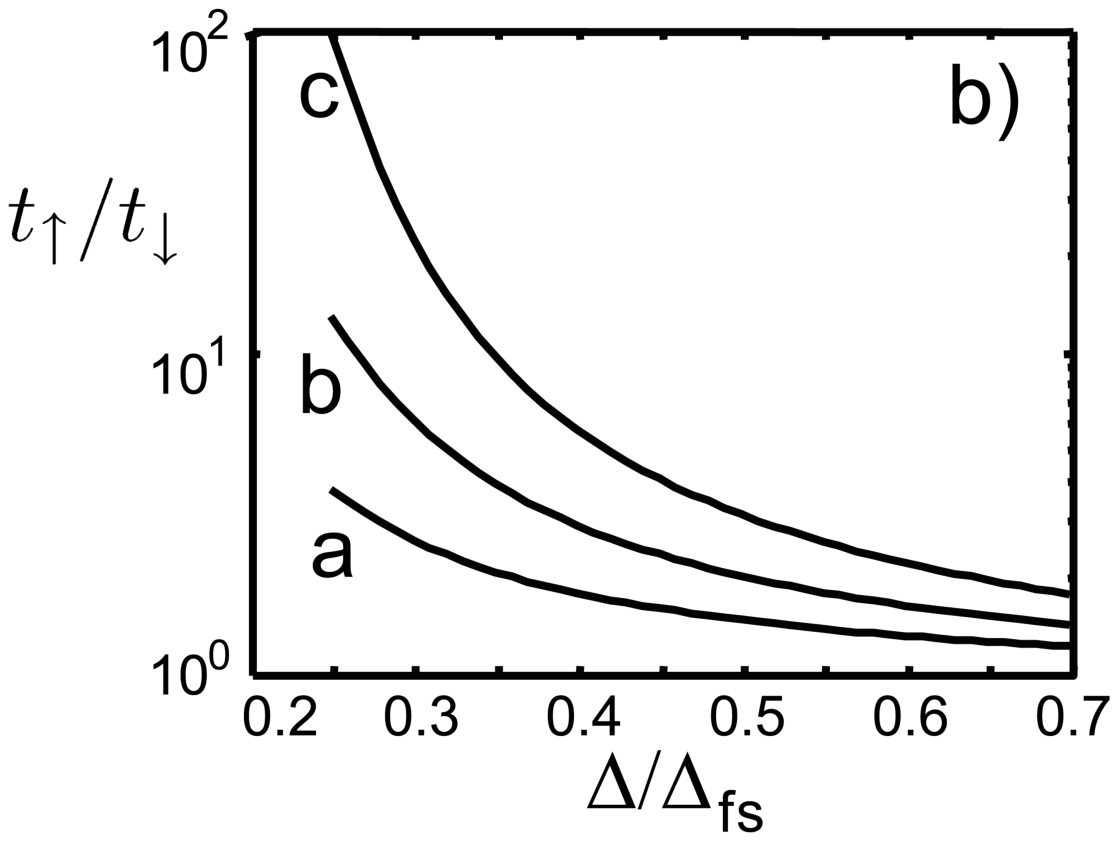}
\end{center}
%% \begin{tabular}{c}
%% \includegraphics[width=\linewidth]{fig4a}\\
%% \includegraphics[width=\linewidth]{fig4b}
%% \end{tabular}
\caption{a) AC Starkshift (in arbitrary units) of the $M_{F}=9/2$
   (solid line) and the $M_{F}=7/2$ state (dashed line) in the
   detuning region right of the interference zero $%
   0.248<\Delta /\Delta _{\mathrm{fs}}$ of the $M_{F}=7/2$ state. b)
   Ratio of the hopping matrix elements for the $t_{\uparrow\equiv
   |M_F=9/2\rangle, \ell }/t_{\downarrow\equiv |M_F=7/2\rangle, \ell
   }$ for a given spatial direction $\ell=1,2,3$ as a function of
   detuning $\Delta$ for a fixed $V_{{\uparrow \equiv
   {|M_{F}{=9/2}\rangle}},\ell }^{(0)}\equiv V_{0}$ with $V_0=5$
   (curve a),$10$ (curve b), and $20 E_R$ (curve c) in units of the
   recoil energy $E_R=\hbar^2 k^{ 2}/2m$. The hopping matrix elements
   were obtained from a band structure calculation for the given depth
   of the optical potential.} \label{fig:pz4}
\end{figure}

This shows how to create a 1D spin dependent lattice
in counterpropagating laser beams of unbalanced $\sigma^{\pm }$
polarization. The set-up is readily generalized to higher dimensions.
For example, a 3D lattice is obtained by first applying a magnetic
field to provide a Zeeman splitting of the hyperfine ground states
(to suppress the cross AC Stark terms), and then applying three
standing wave $\sigma ^{+}$ polarized beams which are tilted by
$45^{0}$ relative to the magnetic field. Figs.~\ref{fig:pz3} and
\ref{fig:pz4} summarize the corresponding results for the $M_{F}=9/2$ and $%
7/2$ hyperfine structure ground states (compare the level scheme
in Fig.~\ref{fig:pz2}b). For these states the squares of the
relevant Clebsch-Gordan coefficients (as defined above) are given
by $|a|^{2}=1/9$
and $|b|^{2}=8/9$. Fig.~\ref{fig:pz3} plots the AC Stark shift of the $%
M_{F}=9/2$ (solid line) and $M_{F}=7/2$ states (dashed line) as a
function of the laser detuning $\Delta $. The detuning interval
covers the region of the $^{2}$P$_{1/2}$ and $^{2}$P$_{3/2}$
excited fine structure states, which are separated by $\Delta
_{\mathrm{fs}}$. We see that for a detuning $\Delta \approx
0.248~\Delta _{\mathrm{fs}}$ the AC Starkshift of the $M_{F}=7/2$
state has a zero while the shift of the $M_{F}=7/2$ varies
comparatively slowly. For detunings in the range $0.248<\Delta
/\Delta _{\mathrm{fs}}<1$ the AC Stark shift of both states
$M_{F}=9/2,7/2$ is negative, i.e. the minima of the optical wells
$V_{\sigma }(\Bx)$ coincide but the wells have different
depths giving rise to a spin-dependent tunneling. The AC
Starkshifts in this region between the two fine structure states
are plotted in Fig.~\ref{fig:pz4}a where again the $M_{F}=9/2$
level is represented by the solid line, and the $M_{F}=7/2$ level
of $^{40}$K is given by the dashed line. In Fig.~\ref{fig:pz4}b we
give the ratio of the corresponding hopping matrix elements as a
function of $\Delta $ in the same interval. Here we have  chosen
a light intensity so that we keep the depth of the optical
potential $V_{{\uparrow \equiv{|M_{F}{=9/2}\rangle}},\ell }^{(0)}\equiv
V_{0}$ (for a given direction $\ell =1,2,3$) at a fixed given
value $V_{0}$
for the whole range of detunings, while $V_{{\downarrow \equiv
{|M_{F}{=7/2}\rangle}}, \ell }^{(0)}$  varies.  Thus the ratio of
$t_{\uparrow \ell}/t_{\downarrow \ell }$ varies.
From Fig.~\ref{fig:pz4}b we see that the
ratio of the tunneling elements can be varied by more than an
order of magnitude when we approach the interference minimum at
$\Delta \approx 0.248~\Delta _{\mathrm{fs}}$.

Besides the alkali atoms discussed above, experimental progress
might soon allow the realization of quantum degenerate fermi gases
with alkaline earth atoms, e.g., with Sr atoms
~\cite{Mukaiyama+KatoriETAL-LaserCoolingSrFermi:03,Xu+YeETAL-LaserCoolingAlkalineEarth:03}.
Alkaline earth atoms have besides their singlet ground states
longlived electronic excited triplet states. Applying an
off-resonant laser field, these states will in general have
quite different AC Stark shifts
~\cite{Katori+TakamotoETAL-OpticalPotentialAtomicClock:03}.
Identifying the ground and metastable excited state with the
spin-up and spin-down states, we thus can also have a natural realization
of a Hubbard model with a spin-dependent interaction.

\subsection{Remarks on spin superposition by Rabi coupling}
\label{sec:Omega>0}

The model we have been able (conceptually) to engineer contains a
Rabi coupling that is not present in the conventional Hubbard
model (\ref{eq:HubbardH:atom}). A few remarks about this issue are
in order.

$\Omega_0=0$ corresponds to the case that
the particle number in each spin species is
conserved separately. For $\Omega_0\neq 0$
only total particle
number conserved, while the relative particle
number can vary.  This point will be important
for our discussion of exotic phases.

To treat the effect of the Rabi term theoretically,
we should first diagonalize the quadratic part of the
Hamiltonian.
This is best done in momentum space. In principle, we
can always find a unitary transformation of the original fermion fields,
\be
\td{c}_\sig(\Bk) =\SU_{\sig\sig^\prime}(\Bk) \, c_\sig(\Bk) \,,
\label{eq:c_trans}
\ee
that transforms the Hamiltonian into
\bea
\td{H} &=& \sum_{\Bk\sig} \td{\eps}_{\sig\Bk} \td{c}^\dag_{\sig \Bk}
   \td{c}_{\sig \Bk}    \label{eq:Htilde} \\
&& -  \sum_{\{k_i,\sig_i\}}
   V_{\sig_1..\sig_4; \Bk_1..\Bk_4}
   \td{c}^\dag_{\sig_1\Bk_1} \td{c}^\dag_{\sig_2\Bk_2}
   \td{c}_{\sig_3\Bk_3} \td{c}_{\sig_4\Bk_4}  \nn
\eea
with
$V_{\sig_1..\sig_4; \Bk_1..\Bk_4} = U\, \del^d(\Bk_1+\Bk_2-\Bk_3-\Bk_4)\,
\SU_{\sig_1\up}(\Bk_1)  \times  \SU_{\sig_2\down}(\Bk_2)
\SU^\dag_{\sig_3\down}(\Bk_3)\SU^\dag_{\sig_4\up}(\Bk_4)\,$ in
$d$-spacial dimensions.
The explicit form of the unitary transformation matrix $\SU(\Bk)$ for each
mode $\Bk$ can be found easily after some elementary algebra, but is not
essential to our discussion here.
The important point is that the new, effective energy bands
$\td{\eps}_{\sig\Bk}$ remain spin dependent. The free parameters
$h$, $\Omega_0$, and $t_\sig$'s allow great freedom of tuning the
new band mass ratio and the Fermi surface difference of the effective up and
down fermions.
Instead of working with the Hamiltonian (\ref{eq:HubbardH:atom}) of the
original fermions, we could just as well start with the Hamiltonian
(\ref{eq:Htilde}) of the effective fermions. The new interaction term will
contain not only terms $s$-wave spin-singlet terms, but also other higher
angular and spin terms.  As far as our main
interest, $s$-wave spin-singlet pairing, is
concerned, the theoretical treatment does not differ substantially
from the model (\ref{eq:HubbardH}) below, for which the Rabi term is
transformed away.

\section{BP to BCS Transition}

Recently there has been considerable discussion
of the possible existence of homogeneous
zero-temperature phases
wherein a superfluid condensation coexists with
one or more Fermi surfaces, where the gap
vanishes. We have
in mind that these are full-fledged codimension 1
Fermi surfaces (e.g., two-dimensional surfaces,
which bound three-dimensional
regions, in a
three-dimensional system) where the gap vanishes.
The roots of this idea go back to early work of
Sarma~\cite{Sarma:63}; the issue has arisen again,
under several
different names, in
several new contexts, including high-density QCD
and (as emphasized here) cold atom systems.
There are delicate issues of stability involved,
which have been mishandled in much of the
literature.
We believe that a careful and correct discussion
is supplied in Ref.~\cite{Forbes:04}.
We shall not repeat the analysis given there, but
we rely on the conclusion: there are a number of
physically
interesting circumstances in which two-component
interacting Fermi systems, of the general type
discussed
in the preceding section, can support a
continuous ``breached pairing'' (BP) to BCS 
quantum phase transition (i.e., a ``superfluid +
normal$\rightarrow$superfluid'' transition at zero temperature).

Stability of the BP phase, which has coexisting, homogeneous
superfluid and normal components at zero-temperature, 
appears to require momentum-dependent
interactions, same-species repulsion, or some
other complication not present in the simplest
Hubbard
Hamiltonian with two states defining a quasi-spin degree of freedom.
For the sake of
simplicity and
concreteness, however, we shall work with this
Hamiltonian.  The qualitative, universal features
of the phase
transition ought not to depend on this idealization. Then,
\begin{eqnarray}
H & =&-\sum_{\sigma\langle i,j\rangle} t_{\sigma}\left(  c_{\sigma
i}^{\dagger }c_{\sigma j}+\mathrm{h.c.}\right)
  +h \sum_{i}\left(c_{\uparrow i}^{\dagger}c_{\uparrow i}
- c_{\down i}^{\dagger}c_{\down i}  \right)  \nn \\
&& -U\sum_{i}c_{\uparrow i}^{\dagger}c_{\downarrow
i}^{\dagger}c_{\downarrow i}c_{\uparrow i}\,.
\label{eq:HubbardH}   %%\label{HamilHubbardlattice}
\end{eqnarray}
where as usual, $c_{\sig i}$ and $c^\dag_{\sig i}$ are fermion
annihilation and creation operators at site $i$,
with $\sig=\up,\down$ indicating
two internal quantum states. 

The Hamiltonian (\ref{eq:HubbardH})
appears to be a
simplified version of (\ref{eq:HubbardH:atom}), by assuming that the
Rabi coupling term is transformed away through
(\ref{eq:c_trans}). An important difference is that the spin up and down
internal states are coherent now.
%In other words, only the total
%particle number of both spins is conserved.
The two quasi-spin states
could well be two hyperfine
spin levels for the case of cold atoms, as we
discussed in detail above. 

\subsection{Nature of quantum phase transition between BCS  and BP
states}
\label{sec:QFT-nature}

\begin{figure}[htbp]
\begin{center}
\includegraphics[width=\linewidth]{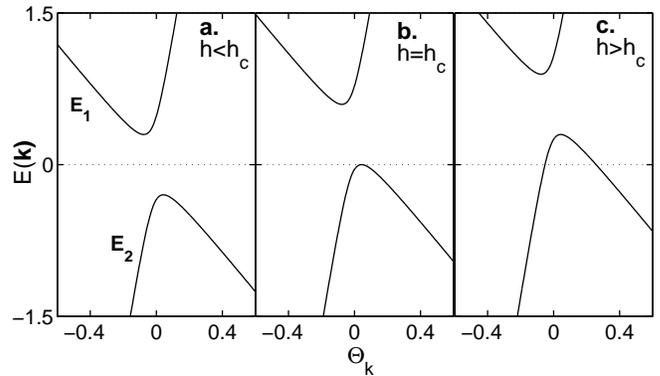}
\end{center}
\caption{Quantum phase transition seen from change of quasiparticle
   spectra. The critical value $h=h_c$ is determined by requiring that
   $E_2$ branch be of just one solution at zero energy. For $h>h_c$,
   one should imagine a three dimensional momentum space in which two
   surfaces, defined by $\Theta_\Bk=\Theta^-$ and
   $\Theta_\Bk=\Theta^+$, are gapless for the $E_2$ branch. a) BCS
   superfluid; b) Quantum critical state; c) ``superfluid+normal'' 
   BP state.  For convenience, $t_\up \geq t_\down$ is always
   assumed throughout the paper.}
\label{fig:cr_qp}
\end{figure}

The general nature of the transition can be gleaned from Fig.~\ref{fig:cr_qp}.
The BP to BCS transition
is continuous, at least in mean field theory. To
characterize its critical dynamics, we must use
paramaeters that
characterize
he singular behavior --- in particular, the
appearance or disappearance of low-energy degrees
of freedom.  The magnitude of
the superconducting order
parameter is evidently not a suitable order
parameter, since it evolves smoothly  (at the
level of mean field theory) and is non-vanishing at both sides
of the transition.
Furthermore, there is no evident change in symmetry
at the transition.
We propose that a good order parameter for the BP
to BCS transition is the quasi-spin polarization
(see
Section~\ref{sec:M}).

\subsection{Quasiparticle excitations}  \label{sec:qp}

We shall assume a uniform pairing,
with order parameter,
$\Del= U\avg{c^\dag_{i\down}c^\dag_{i\up}}
=\mathrm{const.}$
To find the spectrum of low energy quasiparticle
excitations, we diagonalize the mean field Hamiltonian
\be
H_\mathrm{m} = \sum_\Bk \Psi^\dag_\Bk
\pmatrix{2t_\up\Theta_\Bk+h-\mu  & \Del
\cr
\Del^*  & -(2t_\down \Theta_\Bk-h) +\mu }  \Psi_\Bk \,,   \label{eq:Hm}
\ee
where
$$ \Psi_\Bk \equiv  \pmatrix{c_{\Bk\up} \cr c^\dag_{-\Bk\down}} $$
and
$$\Theta_\Bk = -\sum_{l=1..d} \cos(k_l a)\,.$$
(Here $d=3$ for three
dimensional space.)
There are two physical branches of (Bogoliubov) quasiparticle
excitations that have distinct property in a BP state. They are
\be
E_{1,2}(\Bk) = (h+ t_-\Theta_\Bk) \pm \sqrt{(t_+\Theta_\Bk -\mu)^2
   +|\Del|^2}     \label{eq:E12}
\ee
where we have adopted the short-hand notation $t_\pm \equiv t_\up \pm
t_\down$. We will assume $t_\up \geq t_\down$ and $h>h_\times\equiv -(t_-/t_+)
\mu$.   By definition, $h_\times$ is the
value of the magnetic field for which the spin up and down bands
have the same Fermi surface.

Fig.~\ref{fig:cr_qp} shows how the two branches of excitations evolve
with $h$. A quantum phase transition takes place at the critical value
of $h=h_c$ with \cite{remark:hc}
\be
h_c = {-\mu t_- + 2|\Del| \sqrt{t_\up t_\down} \over t_+} \,.
\ee
An important feature of the BP state is that the $E_2$ branch of
excitations crosses the zero energy axis at two
two-dimensional surfaces in three-dimensional
momentum space, defined to be $\Theta_\Bk=\Theta^-$ and
$\Theta_\Bk=\Theta^+$.  Their values are
\be
\Theta^{\pm} = {(\mu t_+ +h t_-) \pm \sqrt{(h t_+ +\mu t_-)^2 -4
     |\Del|^2 t_\up t_\down}  \over 4 t_\up t_\down}\,.   \label{eq:Theta+-}
\ee
At the critical point $h=h_c$,
the two gapless ``Fermi'' surfaces merge into a single one,
\be
\Theta^-=\Theta^+=
\Theta_c \equiv {\mu t_+ + h_c t_- \over 4 t_\up t_\down} \,. \label{eq:TTTc}
\ee

Excitation spectra similar to Fig.~\ref{fig:cr_qp}(c) are also found
in ferromagnetic metals.  It was argued in
~\cite{Karchev+:01} that superconductivity could coexist with
ferromagnetism.

\subsection{As a Lifshitz topological transition}

Lifshitz (topological) transitions \cite{Lifshitz:60} can take
place in metals and alloys at the motion of von Hove singularities
of the electron density of states across Fermi surfaces (for
review, see, e.g., Blanter {\it et al.}~\cite{Blanter:94}).  The
transition from BCS to BP actually falls into this universality
class.
% unlike other magnetic quantum phase transitions studied in
% a variety of condensed matter systems (such as heavy fermion
% compounds).

To illustrate the topological nature of the transition, we start with
the many-body wavefunctions for the BP and BCS states,
\bea
|\Psi_{BCS}\> &=& \prod_\Bk (u_\Bk + v_\Bk c^\dag_{\Bk\up}
c^\dag_{-\Bk\down}) |0\> \,,
\\
|\Psi_{BP}\> &=&\prod_{\Bk: \Theta_\Bk<\Theta^-} (u_\Bk + v_\Bk c^\dag_{\Bk\up}
c^\dag_{-\Bk\down}) \prod_{\Bk: \Theta_\Bk\in [\Theta^-, \Theta^+]}
c^\dag_{\Bk\down} \nn \\
&& \times
\prod_{\Bk: \Theta_\Bk>\Theta^+} (u_\Bk + v_\Bk c^\dag_{\Bk\up}
c^\dag_{-\Bk\down})
  |0\> \,.  \label{eq:PsiBP}
\eea
Here, $u_\Bk$ and $v_\Bk$ are complex numbers satisfying $|u_\Bk|^2  +
|v_\Bk|^2 =1$; their amplitudes are determined by diagonalizing the
mean field Hamiltonian (\ref{eq:Hm}),
%(Bogoliubov) transforming the fermion operators,
%\be
%N^\prime_\Bk = \pmatrix{u_\Bp & v_\Bp  \cr
%       -v_\Bp  & u_\Bp} N_\Bk
%\ee
%with
\be\left\{
\begin{array}{c}
|u_\Bk|^2 \\ |v_\Bk|^2
\end{array}\right\}
  =  {1\over 2} \left( 1 \pm {t_+\Theta_\Bk -\mu  \over
\sqrt{[t_+\Theta_\Bk -\mu]^2 + |\Del|^2 }}\right) \,.
\ee
There is a manifold
of degenerate states featuring an overall relative phase between
the $u_\Bk$ and $v_\Bk$ factors, corresponding to the broken \Uch.
(Note that by our convention of $h>h_\times$,
there are more particles in the spin $\down$ species than the
$\up$ species.

The occupation numbers of both spin up and down
fermions are readily determined.
For the BCS state,
\be
n_{\Bk\up}=n_{\Bk\down} =|v_\Bk|^2\,, \qquad \mbox{(for all $\Bk$)}\,.
\ee
while for the BP state:
%% \be
%% \begin{array}{rcl}
%% \Theta_\Bk<\Theta^- :
%% & \quad & n_{\Bk\up}=n_{\Bk\down} = |v_\Bk|^2 \,,
%% \\
%% {\Theta_\Bk\in[\Theta^-,\Theta^+] :} && n_{\Bk\up}=0\,, \quad
%% n_{\Bk\down} =1\,,\\
%% \Theta_\Bk>\Theta^+ :
%% & \quad & n_{\Bk\up}=n_{\Bk\down} = |v_\Bk|^2 \,,
%% \end{array}
%% \ee
\be
\begin{array}{rcl}
\mbox{if $\Theta_\Bk\in[\Theta^-,\Theta^+]$:} && n_{\Bk\up}=0\,, \quad
n_{\Bk\down} =1\,,\\
\mbox{otherwise:} & \quad & n_{\Bk\up}=n_{\Bk\down} = |v_\Bk|^2 \,.
\label{eq:n(k)}
\end{array}
\ee
In the BCS state the occupation numbers are equal,
while for the BP state they differ.  Since the BCS
state is maintained for a finite range of
parameters, its continuous evolution to the BP
state is non-analytic.
This signals a (zero-temperature, quantum) phase transition.

\begin{figure}[htbp]
\begin{center}
\includegraphics[width=\linewidth]{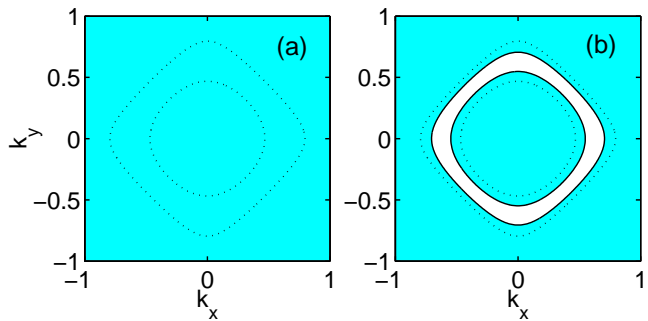}
\end{center}
\caption{An illustration of changing topology of the spin up Fermi
   sea. Shown are cross-sections at $k_z=\pi/2a$ in a 3D first
   Brillouin zone for (a) BCS and (b) BP; momenta units for $k_x, k_y$
   are $\pi/a$ where $a$ is the cubic lattice constant. As usual, the
   occupied states spread out over the entire Brillouin zone due to
   pairing interaction.  The shading in the graph indicates that the
   region has a non-vanishing occupation number. The dotted lines
   indicate the orginal free Fermi surfaces of spin ups (inner curve)
   and spin downs (outer curve). The solid lines in (b) are energy
   surfaces defined by $\Theta_\Bk=\Theta^-$ and
   $\Theta_\Bk=\Theta^+$; For small $\Del\rightarrow 0$, they merge
   with the free Fermi surfaces (dotted lines), respectively. The
   phase transition occurs at the point in which a simply connected
   Fermi sea is isolated into two regions.}
\label{fig:Fermi_sea}
\end{figure}

The phase transition  associated with a change in topology of the Fermi sea.
This kind of transition is
generally known as Lifshitz transition. As shown in
Fig.~\ref{fig:Fermi_sea}, the quantum phase transition connects states of
a simply connected Fermi sea and of isolated two regions.

A transition between superfluid phases of different topology in
momentum space is also known to occur in, for instance, $^3$He-B
phase. There, the superfluid velocity with respect to a container
drives a quantum phase transition from a fully gapped state to
a state of (gapless) fermi surface~\cite{Vollhardt:80,Volovik:03bk}.

\subsection{The order parameter} \label{sec:M}

We now come to discuss the order parameter that characterizes
the quantum phase transition. It is  spin
polarization, $M\equiv \avg{S_z(\Bx)}=  n_\up - n_\down$. However,
unlike a usual Landau-Ginzburg type
theory, no obvious spontaneously broken symmetry is
involved at the transition.
The spin polarization density is given by
\be M = -\int^{\Theta^+}_{\Theta^-} d\Theta\, N(\Theta)  \,,
\ee
where
$$N(\Theta) \equiv  \sum_\Bk
\del(\Theta-\Theta_\Bk)
$$
plays the role of (dimensionless) density of states.
Near the critical point $h=h_c$, both $\Theta^-$ and $\Theta^+$
approach $\Theta_c$. So the spin polarization may be approximated to
the lowest order in $h-h_c$ by
\bea
M &\simeq& N(\Theta_c) (\Theta^--\Theta^+) \nn \\
& =& - N(\Theta_c) {
\sqrt{ (\mu \,{t_-} + h\,{t_+})^2 - 4{\Delta }^2 {t_\down}\,{t_\up}}
\over 2 t_\up t_\down} \,,   \label{eq:M=} \\
&=& - N(\Theta_c) {t_+\over 2 t_\up t_\down} \sqrt{(h-h_c)
   (h+h_c-2h_\times)} \,. \nn
\eea

The spin polarization may be thought being derived from the grand
thermodynamical potential $\Omega$  by viewing it
as a functional of $h$:
$$
M= {\del \Omega[h] \over \del h} \,.
$$
This formula can be inverted. Define $h_M$ as the magnetic field   for
which the above  equation has a prescribed  value $M$.
Then the quantum effective potential $\Gamma[M]$ is defined (as a
functional of $M$, not $h$) by the Legendre transformation
$$
\Gamma[M] = - h M + \Omega[h]\,,
$$
with
  $$
h=h_M 
$$
fixed.
The form of $\Gamma[M]$ is determined by requiring that the equation of
state,
\be
{\pt \Gamma[M] \over \pt M} = -h_M\,,
\ee
produce what is equivalent to Eq.~(\ref{eq:M=}) upon inverting
the latter.   This leads to
\bea
\Gamma[M] &=&  {t_-\over t_+}\mu M +  {\Del\sqrt{t_\up t_\down} \over t_+}
      \left[{|M|}\,\sqrt{1 + \left({M\over
  M_\Del}\right)^2}  \right. \nn \\
&& \qquad\qquad \left. +  M_\Del
     \mathrm{arcsinh}\left({|M|\over M_\Del}\right) \right]  \label{eq:V(M)}
\eea
where  $M_\Del$ is a constant,
$M_\Del \equiv \Del N(\Theta_c)/\sqrt{t_\up t_\down}$. In
deriving Eq.~(\ref{eq:V(M)}),  we
have treated $\Del$ as a magnetic field-independent parameter to
effectively represent the coupling strength. The effective potential
is not analytic in $M$; the appearance of $|M|$  is a consequence of
selecting the physically stable solution of it.
The physical origin is
due to the presence of two gapless ``Fermi'' surfaces.
For small $M$ (and also $M<0$ in our
particular case of $t_\up > t_\down$ and $h>h_\times$),
we could expand $\Gamma[M]$ in
powers of $M$. Then,  in the presence of external magnetic field $h$,
the final effective potential reads
\bea
V(M) &\equiv & h M + \Gamma[M] \nn \\
&=& (h-h_c) M   + { \Del\sqrt{t_\up t_\down} \over 3 t_+ M_\Del^2 } |M|^3
  +\cdots  \label{eq:Vh(M)}
\eea
For $h>h_c$, $V(M)$ always has a minimum at a nonzero $M$.
From either Eq.~(\ref{eq:M=}) or (\ref{eq:Vh(M)}), one can immediately
verify the following power-law
(scaling) relation for the spin polarization at mean field level (for
$h$ near $h_c$)
\be
M =\left\{\begin{array}{ll}
  - {N(\Theta_c) (t_+ \Del)^{1\over 2} \over (t_\up t_\down)^{3\over
      4}}  (h-h_c)^{1/2}\,, & \mbox{if $h>h_c$}\,,
\\
0\,, & \mbox{otherwise} \,.
\end{array} \right.  \label{eq:Mscaling}
\ee

\begin{figure}[htbp]
\begin{center}
\includegraphics[width=\linewidth]{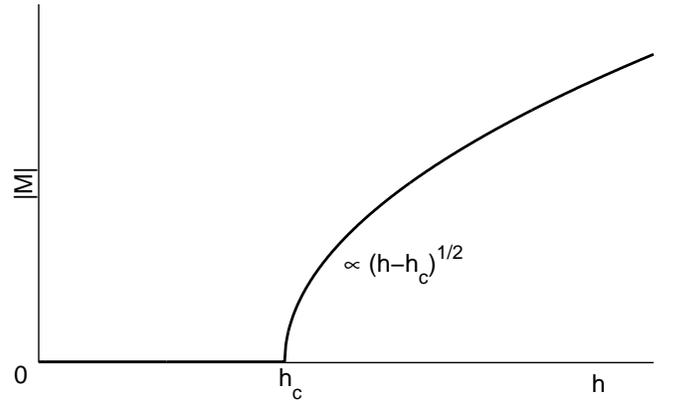}
\end{center}
\caption{Quantum phase transition shown by spin polarization $M$. For
  the BCS (BP) state, $M=0$ ($M\neq 0$).}
\label{fig:cr_M}
\end{figure}

The results are  sketched in Fig.~\ref{fig:cr_M}.
The behavior
of spin polarization near the point $h=h_c$ suggests that
the quantum phase transition is of second order.

\subsection{Scaling theory of $z=2$ gapless fermions}

At the critical point, the $E_2(\Bk)$ branch of quasiparticle excitations
is gapless at a single `Fermi surface' defined by $\Theta(\Bk) =\Theta_c$. What
makes the critical point special is that
the spectrum  disperses quadratically near the surface:
\be
-E_2(\Bk) = {\Bl^2\over 2 m^*_2 }\,, \qquad
  \Bk \equiv \BK+\Bl  \label{eq:E2:K+l}
\ee
where $\BK$ is a vector on the critical Fermi surface and $\Bl$ a
vector of (small) fluctuating momentum orthogonal to the Fermi
surface.
$m^*_2$ is effective band mass that is implicitly determined from
Eq.~(\ref{eq:E12}).
The dispersion relation implies that
the dynamical exponent is  $z=2$ (at least at tree-graph level) in
contrast with $z=1$ in a nominal (metallic) Fermi liquid
\cite{Shankar:94}.

The critical properties of the $E_2(\Bk)$ fermions appear to define a new
universality class.  To begin understanding it,
let us consider the relevant operators.  Consider
a
theory of gapless fermions described by an effective
free action plus possible quartic interactions,
\begin{widetext}
\bea
I_\psi &= & \int dt d^{d-1}\BK d\Bl \quad
\psi^\dag(\Bk) \left(i\pt_t -{\Bl^2\over
   2m^*_2}\right)\psi(\Bk) \nn \\
&& + g\int dt \prod_{j=1..4}\left[ d^{d-1}\BK_j d\Bl_j\right]
  \psi^\dag(\Bk_1)\psi^\dag(\Bk_2) \psi(\Bk_3)\psi(\Bk_4)
  \del^d(\Bk_1+\Bk_2-\Bk_3-\Bk_4) \nn \\
&& + \psi\psi\psi\psi + \psi^\dag \psi\psi\psi\ldots
\eea
where $\psi$ is the gapless fermion field.
\end{widetext}
One can perform a simple renormalization group analysis to find out
how the interactions should scale. When we scale down the fluctuating
momentum $\Bl$ by a factor $s<1$,
$$
\Bl \rightarrow s\Bl \,,
$$
the quasiparticle energy scales down with a dynamical exponent $z=2$
as
$$
\omega \rightarrow s^2 \omega
$$
for a quadratic dispersion relationship.  So, in geneneral, the
momentum and time in the above action should scale as follows:
\bea
&& dt\rightarrow s^{-2} dt\,, \quad
d^{d-1}\BK \rightarrow s^0 d^{d-1} \BK \,, \quad
d\Bl \rightarrow s^1 d\Bl \,, \nn \\
&&
\partial_t\rightarrow s^{2} \partial_t\,, \quad
\Bl \rightarrow s\Bl\,, \qquad \mbox{with $s\rightarrow 0$.}\nn
\eea
Note that the momentum $\BK$, being attached onto the critical Fermi
surface, does not scale.
Accordingly, the fermion field scales as
$$
\psi \rightarrow s^{1/2} \psi \,.
$$

The scaling dimension of potential interaction operators can now be derived
by straightforward power counting.
We find that
a \emph{generic} four-fermion scattering operator is marginal.
That is,
$$
g \rightarrow \left\{
\begin{array}{ll}
s^0 g\,, & \mbox{(generic scatterings)} \\
s^{-1} g \,, & \mbox{(BCS or forward scatterings)} \,.
\end{array}\right.
$$
This result contrasts dramatically with what is found in the standard
renormalization group theory of a conventional Fermi liquid (see, for
instance, the review by
Shankar~\cite{Shankar:94}).  In that context,
only special momentum
configurations are marginal.

%%%%%%%%%%%GOT TO HERE

We next turn to the scaling property of the order parameter
correlation function. We now attempt to derive the
critical theory of quantum fluctuations of
the order parameter field $M$. While the static part of it is already
derived in Eq.~(\ref{eq:Vh(M)}), its dynamical part can be obtained by
calculating the spin-spin correlation,
which is just the fermion polarization bubble diagram, i.e.,
$\avg{S_z(\Bk, i\omega) S_z(-\Bk, -i\omega)} = \Pi(\Bk, i\omega)$ (in
imaginary time formalism), where $S_z(\Bk, i\omega)$ is defined as the
Fourier tansform of 
$$S_{zi}={1\over 2} \left( c^\dag_{\up i} c_{\up i} -
c^\dag_{\down i} c_{\down i}\right)
$$ 
at site $i$ and time $t$. 
The quadratic part of the action for $M$ is
given by $\Pi(\Bk,i\omega)$. Near the critical point, we
find the quantum effective action of $M$ takes the form,
\bea
S[M] &=&  \int d^d x d\tau \Big[ {c_t\over 2}
         (\partial_\tau M)^2 + {c_s\over 2} (\nabla M)^2 \nn \\
  &&
        + {r \over 2} M^2 +{g \over 3! } |M|^3 + \cdots \Big]
\label{eq:S[M]}
\eea
where
\be
r = {2 (t_\up t_\down)^{3\over 4} \over   N(\Theta_c) (t_+
   \Del)^{1\over 2} } (h-h_c)^{1\over 2} \,, \quad
g = {2 (t_\up t_\down)^{3\over 2} \over N^2(\Theta_c) t_+ \Del}  \,,
\ee
and
\be
c_t = {\pt \Pi(\Bk, i\omega) \over \pt(\omega^2)}\,, \qquad
c_s = {{\pt \Pi}(\Bk, i\omega) \over \pt(\Bk^2)}\,,
\ee
with $k,\omega\rightarrow 0$ (long wavelength limit). In deriving the
effective action, we have shifted the position of the minima of
$V(M)$ to eliminate the linear term of Eq.~(\ref{eq:Vh(M)}).

One recognizes that the effective theory of spin polarization field
$M$ bears superficial resemblance to Fisher's~\cite{Fisher:78} $i\phi^3$
famous field theory of the Yang-Lee edge singularity in a
ferromagnetic Ising model. However, the two theories differ
fundamentally in that the cubic interaction term here, $|M|^3$,  is both
non-analytic and real (as opposed to analytic and purely imaginary.)

Further investigation of the new critical theory,
including interactions between the low-energy
fermion modes and the
order parameter, is reserved for future work.

\section*{Acknowledgment}

We have benefited from discussions with E. Gubankova, S. Sachdev,
T. Senthil, X.-G. Wen, and especially Michael
Forbes.  This work is supported in part by funds
provided by the U.S. Department of Energy (D.O.E.) under cooperative
research agreement \#DF-FC02-94ER40818. Work at the University of
Innsbruck is supported by the Austrian Science Foundation and EU networks.

%%%%%%%%%%%%%%%

%\bibliographystyle{apsrev}
\bibliography{quantum_phase}

\begin{thebibliography}{47}
\expandafter\ifx\csname natexlab\endcsname\relax\def\natexlab#1{#1}\fi
\expandafter\ifx\csname bibnamefont\endcsname\relax
  \def\bibnamefont#1{#1}\fi
\expandafter\ifx\csname bibfnamefont\endcsname\relax
  \def\bibfnamefont#1{#1}\fi
\expandafter\ifx\csname citenamefont\endcsname\relax
  \def\citenamefont#1{#1}\fi
\expandafter\ifx\csname url\endcsname\relax
  \def\url#1{\texttt{#1}}\fi
\expandafter\ifx\csname urlprefix\endcsname\relax\def\urlprefix{URL }\fi
\providecommand{\bibinfo}[2]{#2}
\providecommand{\eprint}[2][]{\url{#2}}

\bibitem[{UCA()}]{UCA:nature}
\bibinfo{note}{For a review, see {\it Nature} {\bf 416}, 205-246 (2002)}.

\bibitem[{\citenamefont{Cirac and Zoller}(March
  2004)}]{Cirac+Zoller-PhysicsToday:04}
\bibinfo{author}{\bibfnamefont{J.~I.} \bibnamefont{Cirac}} \bibnamefont{and}
  \bibinfo{author}{\bibfnamefont{P.}~\bibnamefont{Zoller}},
  \bibinfo{journal}{Physics Today} pp. \bibinfo{pages}{38--44}
  (\bibinfo{year}{March 2004}).

\bibitem[{\citenamefont{Greiner et~al.}(2002)\citenamefont{Greiner, Mandel,
  Esslinger, H{\"a}nsch, and Bloch}}]{Bloch-SF-Mott:02}
\bibinfo{author}{\bibfnamefont{M.}~\bibnamefont{Greiner}},
  \bibinfo{author}{\bibfnamefont{O.}~\bibnamefont{Mandel}},
  \bibinfo{author}{\bibfnamefont{T.}~\bibnamefont{Esslinger}},
  \bibinfo{author}{\bibfnamefont{T.}~\bibnamefont{H{\"a}nsch}},
  \bibnamefont{and} \bibinfo{author}{\bibfnamefont{I.}~\bibnamefont{Bloch}},
  \bibinfo{journal}{Nature} \textbf{\bibinfo{volume}{415}}, \bibinfo{pages}{39}
  (\bibinfo{year}{2002}).

\bibitem[{\citenamefont{Orzel et~al.}(2001)\citenamefont{Orzel, Tuchman,
  Fenselau, Yasuda, and Kasevich}}]{Orzel+:01}
\bibinfo{author}{\bibfnamefont{C.}~\bibnamefont{Orzel}},
  \bibinfo{author}{\bibfnamefont{A.~K.} \bibnamefont{Tuchman}},
  \bibinfo{author}{\bibfnamefont{M.~L.} \bibnamefont{Fenselau}},
  \bibinfo{author}{\bibfnamefont{M.}~\bibnamefont{Yasuda}}, \bibnamefont{and}
  \bibinfo{author}{\bibfnamefont{M.~A.} \bibnamefont{Kasevich}},
  \bibinfo{journal}{Science} \textbf{\bibinfo{volume}{291}},
  \bibinfo{pages}{2386} (\bibinfo{year}{2001}).

\bibitem[{\citenamefont{Mandel et~al.}(2003)\citenamefont{Mandel, Greiner,
  Widera, Rom, H{\"a}nsch, and Bloch}}]{Bloch-CCEntanglement:03}
\bibinfo{author}{\bibfnamefont{O.}~\bibnamefont{Mandel}},
  \bibinfo{author}{\bibfnamefont{M.}~\bibnamefont{Greiner}},
  \bibinfo{author}{\bibfnamefont{A.}~\bibnamefont{Widera}},
  \bibinfo{author}{\bibfnamefont{T.}~\bibnamefont{Rom}},
  \bibinfo{author}{\bibfnamefont{T.}~\bibnamefont{H{\"a}nsch}},
  \bibnamefont{and} \bibinfo{author}{\bibfnamefont{I.}~\bibnamefont{Bloch}},
  \bibinfo{journal}{Nature} \textbf{\bibinfo{volume}{425}},
  \bibinfo{pages}{937} (\bibinfo{year}{2003}).

\bibitem[{\citenamefont{St\"{o}ferle et~al.}(2004)\citenamefont{St\"{o}ferle,
  Moritz, Schori, K\"ohl, and Esslinger}}]{Stoeferle+Esslinger-Mott1D:04}
\bibinfo{author}{\bibfnamefont{T.}~\bibnamefont{St\"{o}ferle}},
  \bibinfo{author}{\bibfnamefont{H.}~\bibnamefont{Moritz}},
  \bibinfo{author}{\bibfnamefont{C.}~\bibnamefont{Schori}},
  \bibinfo{author}{\bibfnamefont{M.}~\bibnamefont{K\"ohl}}, \bibnamefont{and}
  \bibinfo{author}{\bibfnamefont{T.}~\bibnamefont{Esslinger}},
  \bibinfo{journal}{Phys. Rev. Lett.} \textbf{\bibinfo{volume}{92}},
  \bibinfo{pages}{130403} (\bibinfo{year}{2004}).

\bibitem[{\citenamefont{Jaksch et~al.}(1998)\citenamefont{Jaksch, Bruder,
  Cirac, Gardiner, and
  Zoller}}]{Jaksch+BruderETAL-ColdBosonicAtomOpticalLattices:98}
\bibinfo{author}{\bibfnamefont{D.}~\bibnamefont{Jaksch}},
  \bibinfo{author}{\bibfnamefont{C.}~\bibnamefont{Bruder}},
  \bibinfo{author}{\bibfnamefont{J.}~\bibnamefont{Cirac}},
  \bibinfo{author}{\bibfnamefont{C.}~\bibnamefont{Gardiner}}, \bibnamefont{and}
  \bibinfo{author}{\bibfnamefont{P.}~\bibnamefont{Zoller}},
  \bibinfo{journal}{Phys. Rev. Lett.} \textbf{\bibinfo{volume}{81}},
  \bibinfo{pages}{3108} (\bibinfo{year}{1998}).

\bibitem[{\citenamefont{Jaksch et~al.}(1999)\citenamefont{Jaksch, Briegel,
  Cirac, Gardiner, and Zoller}}]{Jaksch+ETAL-EntanglementColdCollision:99}
\bibinfo{author}{\bibfnamefont{D.}~\bibnamefont{Jaksch}},
  \bibinfo{author}{\bibfnamefont{H.}~\bibnamefont{Briegel}},
  \bibinfo{author}{\bibfnamefont{J.}~\bibnamefont{Cirac}},
  \bibinfo{author}{\bibfnamefont{C.}~\bibnamefont{Gardiner}}, \bibnamefont{and}
  \bibinfo{author}{\bibfnamefont{P.}~\bibnamefont{Zoller}},
  \bibinfo{journal}{Phys. Rev. Lett.} \textbf{\bibinfo{volume}{82}},
  \bibinfo{pages}{1975} (\bibinfo{year}{1999}).

\bibitem[{\citenamefont{Lewenstein et~al.}(2004)\citenamefont{Lewenstein,
  Santos, Baranov, and
  Fehrmann}}]{Lewenstein+SantosETAL-AtomBoseMixtOpticalLattices:04}
\bibinfo{author}{\bibfnamefont{M.}~\bibnamefont{Lewenstein}},
  \bibinfo{author}{\bibfnamefont{L.}~\bibnamefont{Santos}},
  \bibinfo{author}{\bibfnamefont{M.~A.} \bibnamefont{Baranov}},
  \bibnamefont{and} \bibinfo{author}{\bibfnamefont{H.}~\bibnamefont{Fehrmann}},
  \bibinfo{journal}{Phys. Rev. Lett.} \textbf{\bibinfo{volume}{92}},
  \bibinfo{pages}{050401} (\bibinfo{year}{2004}).

\bibitem[{\citenamefont{Damski et~al.}(2003)\citenamefont{Damski, Zakrzewski,
  Santos, Zoller, and Lewenstein}}]{Damski+ZakrzewskiETAL-AtomBoseAndeGlas:03}
\bibinfo{author}{\bibfnamefont{B.}~\bibnamefont{Damski}},
  \bibinfo{author}{\bibfnamefont{J.}~\bibnamefont{Zakrzewski}},
  \bibinfo{author}{\bibfnamefont{L.}~\bibnamefont{Santos}},
  \bibinfo{author}{\bibfnamefont{P.}~\bibnamefont{Zoller}}, \bibnamefont{and}
  \bibinfo{author}{\bibfnamefont{M.}~\bibnamefont{Lewenstein}},
  \bibinfo{journal}{Phys. Rev. Lett.} \textbf{\bibinfo{volume}{91}},
  \bibinfo{pages}{080403} (\bibinfo{year}{2003}).

\bibitem[{Kuk()}]{Kuklov++Duan:03}
\bibinfo{note}{A. B. Kuklov and B. V. Svistunov, Phys. Rev. Lett. {\bf 90},
  100401 (2003); L.-M. Duan, E. Demler and M. D. Lukin, Phys. Rev. Lett. {\bf
  91}, 090402 (2003).}

\bibitem[{\citenamefont{Paredes and
  Cirac}(2003)}]{Paredes+Cirac-CooperPairsLuttingerOpticalLattice:03}
\bibinfo{author}{\bibfnamefont{B.}~\bibnamefont{Paredes}} \bibnamefont{and}
  \bibinfo{author}{\bibfnamefont{J.~I.} \bibnamefont{Cirac}},
  \bibinfo{journal}{Phys. Rev. Lett.} \textbf{\bibinfo{volume}{90}},
  \bibinfo{pages}{150402} (\bibinfo{year}{2003}).

\bibitem[{\citenamefont{Ho et~al.}(2004)\citenamefont{Ho, Cazalilla, and
  Giamarchi}}]{Ho+CazalillaETAL-OptcaliLattice1D2D:04}
\bibinfo{author}{\bibfnamefont{A.~F.} \bibnamefont{Ho}},
  \bibinfo{author}{\bibfnamefont{M.~A.} \bibnamefont{Cazalilla}},
  \bibnamefont{and}
  \bibinfo{author}{\bibfnamefont{T.}~\bibnamefont{Giamarchi}},
  \bibinfo{journal}{Phys. Rev. Lett.} \textbf{\bibinfo{volume}{92}},
  \bibinfo{pages}{130405} (\bibinfo{year}{2004}).

\bibitem[{\citenamefont{Pachos and Rico}()}]{Pachos-Rico:04pre}
\bibinfo{author}{\bibfnamefont{J.~K.} \bibnamefont{Pachos}} \bibnamefont{and}
  \bibinfo{author}{\bibfnamefont{E.}~\bibnamefont{Rico}},
  \bibinfo{note}{quant-ph/0404048}.

\bibitem[{\citenamefont{Hofstetter et~al.}(2002)\citenamefont{Hofstetter,
  Cirac, Zoller, Demler, and Lukin}}]{Hofstetter+:02}
\bibinfo{author}{\bibfnamefont{W.}~\bibnamefont{Hofstetter}},
  \bibinfo{author}{\bibfnamefont{J.}~\bibnamefont{Cirac}},
  \bibinfo{author}{\bibfnamefont{P.}~\bibnamefont{Zoller}},
  \bibinfo{author}{\bibfnamefont{E.}~\bibnamefont{Demler}}, \bibnamefont{and}
  \bibinfo{author}{\bibfnamefont{M.}~\bibnamefont{Lukin}},
  \bibinfo{journal}{Phys. Rev. Lett.} \textbf{\bibinfo{volume}{89}},
  \bibinfo{pages}{220407} (\bibinfo{year}{2002}).

\bibitem[{\citenamefont{Rabl et~al.}(2003)\citenamefont{Rabl, Daley, Fedichev,
  Cirac, and Zoller}}]{Rabl+DaleyETAL-DefectSuppressed:03}
\bibinfo{author}{\bibfnamefont{P.}~\bibnamefont{Rabl}},
  \bibinfo{author}{\bibfnamefont{A.~J.} \bibnamefont{Daley}},
  \bibinfo{author}{\bibfnamefont{P.~O.} \bibnamefont{Fedichev}},
  \bibinfo{author}{\bibfnamefont{J.~I.} \bibnamefont{Cirac}}, \bibnamefont{and}
  \bibinfo{author}{\bibfnamefont{P.}~\bibnamefont{Zoller}},
  \bibinfo{journal}{Phys. Rev. Lett.} \textbf{\bibinfo{volume}{91}},
  \bibinfo{pages}{110403} (\bibinfo{year}{2003}).

\bibitem[{\citenamefont{Strecker et~al.}(2003)\citenamefont{Strecker,
  Partridge, and Hulet}}]{Hulet:03}
\bibinfo{author}{\bibfnamefont{K.~E.} \bibnamefont{Strecker}},
  \bibinfo{author}{\bibfnamefont{G.~B.} \bibnamefont{Partridge}},
  \bibnamefont{and} \bibinfo{author}{\bibfnamefont{R.~G.} \bibnamefont{Hulet}},
  \bibinfo{journal}{Phys. Rev. Lett.} \textbf{\bibinfo{volume}{91}},
  \bibinfo{pages}{080406} (\bibinfo{year}{2003}).

\bibitem[{\citenamefont{Greiner et~al.}(2003)\citenamefont{Greiner, Regal, and
  Jin}}]{Greiner+Regal+Jin:03}
\bibinfo{author}{\bibfnamefont{M.}~\bibnamefont{Greiner}},
  \bibinfo{author}{\bibfnamefont{C.~A.} \bibnamefont{Regal}}, \bibnamefont{and}
  \bibinfo{author}{\bibfnamefont{D.~S.} \bibnamefont{Jin}},
  \bibinfo{journal}{Nature} \textbf{\bibinfo{volume}{426}},
  \bibinfo{pages}{537} (\bibinfo{year}{2003}).

\bibitem[{\citenamefont{Cubizolles et~al.}(2003)\citenamefont{Cubizolles,
  Bourdel, Kokkelmans, Shlyapnikov, and
  Salomon}}]{Cubizolles+Salomon-MolecularBECLi:03}
\bibinfo{author}{\bibfnamefont{J.}~\bibnamefont{Cubizolles}},
  \bibinfo{author}{\bibfnamefont{T.}~\bibnamefont{Bourdel}},
  \bibinfo{author}{\bibfnamefont{S.~J. J. M.~F.} \bibnamefont{Kokkelmans}},
  \bibinfo{author}{\bibfnamefont{G.~V.} \bibnamefont{Shlyapnikov}},
  \bibnamefont{and} \bibinfo{author}{\bibfnamefont{C.}~\bibnamefont{Salomon}},
  \bibinfo{journal}{Phys. Rev. Lett.} \textbf{\bibinfo{volume}{91}},
  \bibinfo{pages}{240401} (\bibinfo{year}{2003}).

\bibitem[{\citenamefont{Zwierlein et~al.}(2003)\citenamefont{Zwierlein, Stan,
  Schunck, Raupach, Gupta, Hadzibabic, and
  Ketterle}}]{Zwierlein+KetterleETAL-LiMolecularBEC:03}
\bibinfo{author}{\bibfnamefont{M.~W.} \bibnamefont{Zwierlein}},
  \bibinfo{author}{\bibfnamefont{C.~A.} \bibnamefont{Stan}},
  \bibinfo{author}{\bibfnamefont{C.~H.} \bibnamefont{Schunck}},
  \bibinfo{author}{\bibfnamefont{S.~M.~F.} \bibnamefont{Raupach}},
  \bibinfo{author}{\bibfnamefont{S.}~\bibnamefont{Gupta}},
  \bibinfo{author}{\bibfnamefont{Z.}~\bibnamefont{Hadzibabic}},
  \bibnamefont{and} \bibinfo{author}{\bibfnamefont{W.}~\bibnamefont{Ketterle}},
  \bibinfo{journal}{Phys. Rev. Lett.} \textbf{\bibinfo{volume}{91}},
  \bibinfo{pages}{250401} (\bibinfo{year}{2003}).

\bibitem[{\citenamefont{Regal et~al.}(2004)\citenamefont{Regal, Greiner, and
  Jin}}]{Regal+Greiner+Jin-KCrossover:04}
\bibinfo{author}{\bibfnamefont{C.~A.} \bibnamefont{Regal}},
  \bibinfo{author}{\bibfnamefont{M.}~\bibnamefont{Greiner}}, \bibnamefont{and}
  \bibinfo{author}{\bibfnamefont{D.~S.} \bibnamefont{Jin}},
  \bibinfo{journal}{Phys. Rev. Lett.} \textbf{\bibinfo{volume}{92}},
  \bibinfo{pages}{040403} (\bibinfo{year}{2004}).

\bibitem[{\citenamefont{Bartenstein et~al.}(2004)\citenamefont{Bartenstein,
  Altmeyer, Riedl, Jochim, Chin, Denschlag, and
  Grimm}}]{Bartenstein+Grimm-LiCrossover:04}
\bibinfo{author}{\bibfnamefont{M.}~\bibnamefont{Bartenstein}},
  \bibinfo{author}{\bibfnamefont{A.}~\bibnamefont{Altmeyer}},
  \bibinfo{author}{\bibfnamefont{S.}~\bibnamefont{Riedl}},
  \bibinfo{author}{\bibfnamefont{S.}~\bibnamefont{Jochim}},
  \bibinfo{author}{\bibfnamefont{C.}~\bibnamefont{Chin}},
  \bibinfo{author}{\bibfnamefont{J.~H.} \bibnamefont{Denschlag}},
  \bibnamefont{and} \bibinfo{author}{\bibfnamefont{R.}~\bibnamefont{Grimm}},
  \bibinfo{journal}{Phys. Rev. Lett.} \textbf{\bibinfo{volume}{92}},
  \bibinfo{pages}{120401} (\bibinfo{year}{2004}).

\bibitem[{\citenamefont{Zwierlein et~al.}(2004)\citenamefont{Zwierlein, Stan,
  Schunck, Raupach, Kerman, and Ketterle}}]{Zwierlein+Ketterle-LiCrossover:04}
\bibinfo{author}{\bibfnamefont{M.~W.} \bibnamefont{Zwierlein}},
  \bibinfo{author}{\bibfnamefont{C.~A.} \bibnamefont{Stan}},
  \bibinfo{author}{\bibfnamefont{C.~H.} \bibnamefont{Schunck}},
  \bibinfo{author}{\bibfnamefont{S.~M.~F.} \bibnamefont{Raupach}},
  \bibinfo{author}{\bibfnamefont{A.~J.} \bibnamefont{Kerman}},
  \bibnamefont{and} \bibinfo{author}{\bibfnamefont{W.}~\bibnamefont{Ketterle}},
  \bibinfo{journal}{Phys. Rev. Lett.} \textbf{\bibinfo{volume}{92}},
  \bibinfo{pages}{120403} (\bibinfo{year}{2004}).

\bibitem[{\citenamefont{Kinast et~al.}()\citenamefont{Kinast, Hemmer, Gehm,
  Turlapov, and Thomas}}]{Thomas:04}
\bibinfo{author}{\bibfnamefont{J.}~\bibnamefont{Kinast}},
  \bibinfo{author}{\bibfnamefont{S.~L.} \bibnamefont{Hemmer}},
  \bibinfo{author}{\bibfnamefont{M.~E.} \bibnamefont{Gehm}},
  \bibinfo{author}{\bibfnamefont{A.}~\bibnamefont{Turlapov}}, \bibnamefont{and}
  \bibinfo{author}{\bibfnamefont{J.~E.} \bibnamefont{Thomas}},
  \bibinfo{note}{cond-mat/0403540}.

\bibitem[{\citenamefont{Ferlaino et~al.}(2004)\citenamefont{Ferlaino,
  de~Mirandes, Roati, Modugno, and
  Inguscio}}]{Ferlaino+Inguscio-ExpansionFermiBose:04}
\bibinfo{author}{\bibfnamefont{F.}~\bibnamefont{Ferlaino}},
  \bibinfo{author}{\bibfnamefont{E.}~\bibnamefont{de~Mirandes}},
  \bibinfo{author}{\bibfnamefont{G.}~\bibnamefont{Roati}},
  \bibinfo{author}{\bibfnamefont{G.}~\bibnamefont{Modugno}}, \bibnamefont{and}
  \bibinfo{author}{\bibfnamefont{M.}~\bibnamefont{Inguscio}},
  \bibinfo{journal}{Phys. Rev. Lett.} \textbf{\bibinfo{volume}{92}},
  \bibinfo{pages}{140405} (\bibinfo{year}{2004}).

\bibitem[{\citenamefont{Liu and Wilczek}(2003)}]{Liu-Wilczek:03}
\bibinfo{author}{\bibfnamefont{W.~V.} \bibnamefont{Liu}} \bibnamefont{and}
  \bibinfo{author}{\bibfnamefont{F.}~\bibnamefont{Wilczek}},
  \bibinfo{journal}{Phys. Rev. Lett.} \textbf{\bibinfo{volume}{90}},
  \bibinfo{pages}{047002} (\bibinfo{year}{2003}).

\bibitem[{\citenamefont{Wu and Yip}(2003)}]{Wu-Yip:03}
\bibinfo{author}{\bibfnamefont{S.-T.} \bibnamefont{Wu}} \bibnamefont{and}
  \bibinfo{author}{\bibfnamefont{S.}~\bibnamefont{Yip}},
  \bibinfo{journal}{Phys. Rev. A} \textbf{\bibinfo{volume}{67}},
  \bibinfo{pages}{053603} (\bibinfo{year}{2003}).

\bibitem[{\citenamefont{Gubankova et~al.}(2003)\citenamefont{Gubankova, Liu,
  and Wilczek}}]{Gubankova+:03}
\bibinfo{author}{\bibfnamefont{E.}~\bibnamefont{Gubankova}},
  \bibinfo{author}{\bibfnamefont{W.~V.} \bibnamefont{Liu}}, \bibnamefont{and}
  \bibinfo{author}{\bibfnamefont{F.}~\bibnamefont{Wilczek}},
  \bibinfo{journal}{Phys. Rev. Lett.} \textbf{\bibinfo{volume}{91}},
  \bibinfo{pages}{032001} (\bibinfo{year}{2003}).

\bibitem[{Liu()}]{Liu-Wilczek:03:comm}
\bibinfo{note}{W. V. Liu and F. Wilczek, cond-mat/0304632}.

\bibitem[{\citenamefont{Bedaque et~al.}(2003)\citenamefont{Bedaque, Caldas, and
  Rupak}}]{Bedaque:03}
\bibinfo{author}{\bibfnamefont{P.~F.} \bibnamefont{Bedaque}},
  \bibinfo{author}{\bibfnamefont{H.}~\bibnamefont{Caldas}}, \bibnamefont{and}
  \bibinfo{author}{\bibfnamefont{G.}~\bibnamefont{Rupak}},
  \bibinfo{journal}{Phys. Rev. Lett.} \textbf{\bibinfo{volume}{91}},
  \bibinfo{pages}{247002} (\bibinfo{year}{2003}).

\bibitem[{\citenamefont{Liao and Zhuang}()}]{Liao:03pre}
\bibinfo{author}{\bibfnamefont{J.}~\bibnamefont{Liao}} \bibnamefont{and}
  \bibinfo{author}{\bibfnamefont{P.}~\bibnamefont{Zhuang}},
  \bibinfo{note}{cond-mat/0307516}.

\bibitem[{\citenamefont{Deb et~al.}()\citenamefont{Deb, Mishra, Mishra, and
  Panigrahi}}]{Deb:03pre}
\bibinfo{author}{\bibfnamefont{B.}~\bibnamefont{Deb}},
  \bibinfo{author}{\bibfnamefont{A.}~\bibnamefont{Mishra}},
  \bibinfo{author}{\bibfnamefont{H.}~\bibnamefont{Mishra}}, \bibnamefont{and}
  \bibinfo{author}{\bibfnamefont{P.~K.} \bibnamefont{Panigrahi}},
  \bibinfo{note}{cond-mat/0308369}.

\bibitem[{\citenamefont{Bardeen et~al.}(1957)\citenamefont{Bardeen, Cooper, and
  Schrieffer}}]{BCS:57}
\bibinfo{author}{\bibfnamefont{J.}~\bibnamefont{Bardeen}},
  \bibinfo{author}{\bibfnamefont{L.~N.} \bibnamefont{Cooper}},
  \bibnamefont{and} \bibinfo{author}{\bibfnamefont{J.~R.}
  \bibnamefont{Schrieffer}}, \bibinfo{journal}{Phys. Rev.}
  \textbf{\bibinfo{volume}{108}}, \bibinfo{pages}{1175} (\bibinfo{year}{1957}).

\bibitem[{\citenamefont{DeMarco and Jin}(1999)}]{Jin:99}
\bibinfo{author}{\bibfnamefont{B.}~\bibnamefont{DeMarco}} \bibnamefont{and}
  \bibinfo{author}{\bibfnamefont{D.~S.} \bibnamefont{Jin}},
  \bibinfo{journal}{Science} \textbf{\bibinfo{volume}{285}},
  \bibinfo{pages}{1703} (\bibinfo{year}{1999}).

\bibitem[{\citenamefont{Mukaiyama et~al.}(2003)\citenamefont{Mukaiyama, Katori,
  Ido, Li, and Kuwata-Gonokami}}]{Mukaiyama+KatoriETAL-LaserCoolingSrFermi:03}
\bibinfo{author}{\bibfnamefont{T.}~\bibnamefont{Mukaiyama}},
  \bibinfo{author}{\bibfnamefont{H.}~\bibnamefont{Katori}},
  \bibinfo{author}{\bibfnamefont{T.}~\bibnamefont{Ido}},
  \bibinfo{author}{\bibfnamefont{Y.}~\bibnamefont{Li}}, \bibnamefont{and}
  \bibinfo{author}{\bibfnamefont{M.}~\bibnamefont{Kuwata-Gonokami}},
  \bibinfo{journal}{Phys. Rev. Lett.} \textbf{\bibinfo{volume}{90}},
  \bibinfo{pages}{113002} (\bibinfo{year}{2003}).

\bibitem[{\citenamefont{Xu et~al.}(2003)\citenamefont{Xu, Loftus, Dunn, Greene,
  Hall, Gallagher, and Ye}}]{Xu+YeETAL-LaserCoolingAlkalineEarth:03}
\bibinfo{author}{\bibfnamefont{X.}~\bibnamefont{Xu}},
  \bibinfo{author}{\bibfnamefont{T.~H.} \bibnamefont{Loftus}},
  \bibinfo{author}{\bibfnamefont{J.~W.} \bibnamefont{Dunn}},
  \bibinfo{author}{\bibfnamefont{C.~H.} \bibnamefont{Greene}},
  \bibinfo{author}{\bibfnamefont{J.~L.} \bibnamefont{Hall}},
  \bibinfo{author}{\bibfnamefont{A.}~\bibnamefont{Gallagher}},
  \bibnamefont{and} \bibinfo{author}{\bibfnamefont{J.}~\bibnamefont{Ye}},
  \bibinfo{journal}{Phys. Rev. Lett.} \textbf{\bibinfo{volume}{90}},
  \bibinfo{pages}{193002} (\bibinfo{year}{2003}).

\bibitem[{\citenamefont{Katori et~al.}(2003)\citenamefont{Katori, Takamoto,
  Pal'chikov, and
  Ovsiannikov}}]{Katori+TakamotoETAL-OpticalPotentialAtomicClock:03}
\bibinfo{author}{\bibfnamefont{H.}~\bibnamefont{Katori}},
  \bibinfo{author}{\bibfnamefont{M.}~\bibnamefont{Takamoto}},
  \bibinfo{author}{\bibfnamefont{V.~G.} \bibnamefont{Pal'chikov}},
  \bibnamefont{and} \bibinfo{author}{\bibfnamefont{V.~D.}
  \bibnamefont{Ovsiannikov}}, \bibinfo{journal}{Phys. Rev. Lett.}
  \textbf{\bibinfo{volume}{91}}, \bibinfo{pages}{173005}
  (\bibinfo{year}{2003}).

\bibitem[{\citenamefont{Sarma}(1963)}]{Sarma:63}
\bibinfo{author}{\bibfnamefont{G.}~\bibnamefont{Sarma}},
  \bibinfo{journal}{Phys. Chem. Solid} \textbf{\bibinfo{volume}{24}},
  \bibinfo{pages}{1029} (\bibinfo{year}{1963}).

\bibitem[{\citenamefont{Forbes et~al.}()\citenamefont{Forbes, Gubankova, Liu,
  and Wilczek}}]{Forbes:04}
\bibinfo{author}{\bibfnamefont{M.~M.} \bibnamefont{Forbes}},
  \bibinfo{author}{\bibfnamefont{E.}~\bibnamefont{Gubankova}},
  \bibinfo{author}{\bibfnamefont{W.~V.} \bibnamefont{Liu}}, \bibnamefont{and}
  \bibinfo{author}{\bibfnamefont{F.}~\bibnamefont{Wilczek}},
  \bibinfo{note}{hep-ph/0405059}.

\bibitem[{rem()}]{remark:hc}
\bibinfo{note}{There is another solution, $h_{c2}$, which also satisfies the
  equation $\Theta^- =\Theta^+$. $h_{c2} < h_\times$ (compared with
  $h_c>h_\times$). For $t_\up\gg t_\down$, $h_{c2}$ is typically negative for a
  chemical potential close to zero; it corresponds to that the heavy fermion
  band has a smaller Fermi surface than the light. That would realize another
  special limit of BP --- one might call ``exterior gap superfluidity".
  However, the essential physics is unaltered.}

\bibitem[{\citenamefont{Karchev et~al.}(2001)\citenamefont{Karchev, Blagoev,
  Bedell, and Littlewood}}]{Karchev+:01}
\bibinfo{author}{\bibfnamefont{N.~I.} \bibnamefont{Karchev}},
  \bibinfo{author}{\bibfnamefont{K.~B.} \bibnamefont{Blagoev}},
  \bibinfo{author}{\bibfnamefont{K.~S.} \bibnamefont{Bedell}},
  \bibnamefont{and} \bibinfo{author}{\bibfnamefont{P.~B.}
  \bibnamefont{Littlewood}}, \bibinfo{journal}{Phys. Rev. Lett.}
  \textbf{\bibinfo{volume}{86}}, \bibinfo{pages}{846} (\bibinfo{year}{2001}),
  \bibinfo{note}{and references therein}.

\bibitem[{\citenamefont{Lifshitz}(1960)}]{Lifshitz:60}
\bibinfo{author}{\bibfnamefont{I.~M.} \bibnamefont{Lifshitz}},
  \bibinfo{journal}{Sov. Phys. JETP} \textbf{\bibinfo{volume}{11}},
  \bibinfo{pages}{1130} (\bibinfo{year}{1960}).

\bibitem[{\citenamefont{Blanter et~al.}(1994)\citenamefont{Blanter, Kaganov,
  Pantsulaya, and Varlamov}}]{Blanter:94}
\bibinfo{author}{\bibfnamefont{Y.~M.} \bibnamefont{Blanter}},
  \bibinfo{author}{\bibfnamefont{M.~I.} \bibnamefont{Kaganov}},
  \bibinfo{author}{\bibfnamefont{A.~V.} \bibnamefont{Pantsulaya}},
  \bibnamefont{and} \bibinfo{author}{\bibfnamefont{A.~A.}
  \bibnamefont{Varlamov}}, \bibinfo{journal}{Phys. Rep.}
  \textbf{\bibinfo{volume}{245}}, \bibinfo{pages}{160} (\bibinfo{year}{1994}).

\bibitem[{\citenamefont{Vollhardt et~al.}(1980)\citenamefont{Vollhardt, Maki,
  and Schopohl}}]{Vollhardt:80}
\bibinfo{author}{\bibfnamefont{D.}~\bibnamefont{Vollhardt}},
  \bibinfo{author}{\bibfnamefont{K.}~\bibnamefont{Maki}}, \bibnamefont{and}
  \bibinfo{author}{\bibfnamefont{N.}~\bibnamefont{Schopohl}},
  \bibinfo{journal}{J. Low Temp. Phys.} \textbf{\bibinfo{volume}{39}},
  \bibinfo{pages}{79} (\bibinfo{year}{1980}).

\bibitem[{\citenamefont{Volovik}(2003)}]{Volovik:03bk}
\bibinfo{author}{\bibfnamefont{G.}~\bibnamefont{Volovik}},
  \emph{\bibinfo{title}{The universe in a Helium droplet}}
  (\bibinfo{publisher}{Oxford University Press}, \bibinfo{address}{New York},
  \bibinfo{year}{2003}), \bibinfo{note}{section 26.1.2}.

\bibitem[{\citenamefont{Shankar}(1994)}]{Shankar:94}
\bibinfo{author}{\bibfnamefont{R.}~\bibnamefont{Shankar}},
  \bibinfo{journal}{Rev. Mod. Phys.} \textbf{\bibinfo{volume}{66}},
  \bibinfo{pages}{129} (\bibinfo{year}{1994}).

\bibitem[{\citenamefont{Fisher}(1978)}]{Fisher:78}
\bibinfo{author}{\bibfnamefont{M.~E.} \bibnamefont{Fisher}},
  \bibinfo{journal}{Phys. Rev. Lett.} \textbf{\bibinfo{volume}{40}},
  \bibinfo{pages}{1610} (\bibinfo{year}{1978}).

\end{thebibliography}

%\begin{thebibliography}{99}
%\end{thebibliography}

%\newpage
%\printfigures
\end{document}